\documentstyle[epsfig]{aipproc}
\begin{document}
\title{Lowest-Lying Scalar Mesons and a Possible Probe of Their Quark
Substructure
\footnote{
Talk given at 21st Annual MRST Conference: High Energy Physics at the
Millenium, Ottawa, Ontario, Canada, 10-12 May 1999.}} 
\author{ Amir H. Fariborz}
\address{ Department of Physics, Syracuse University, Syracuse, New
York 13244-1130, USA.}
\maketitle
\begin{abstract}
In this talk, an overview of the status of the light scalar mesons in the
context of the  non linear chiral Lagrangian of references
\cite{Blk98,Blk99,Far99} is
presented.   The evidence for the existence of a scalar nonet below 1 GeV
is reviewed, and it is shown that by introducing a scalar nonet an
indirect way of probing the quark substructure of these scalars through
the scalar mixing  angle can be obtained.   It is then reviewed that
consistency of this non-linear chiral Lagrangian
framework with the experimental data on $\pi\pi$ and $\pi K$ scattering,
as well as the decay  $\eta'\rightarrow\eta\pi\pi$,  results in a range
for the mixing angle which indicates that the quark substructure of these
light scalars are closer to a four quark picture.
\end{abstract}
\section{Introduction}
Lowest-lying scalar mesons (scalar mesons below 1 GeV) are of fundamental
importance in
understanding the theory and phenomenology of low energy QCD.  
However, the properties of these scalars, in particular their quark
substructure,  are not quite understood.  As a result they are at the
focus  of many theoretical and experimental investigations.

From the experimental point of view,  there are at least four well
established
light scalars --  the isosinglet $f_0(980)$ and the isotriplet $a_0(980)$.
There are also five other candidates --
the isosinglet $\sigma (560)$,  and two
isodoublets $\kappa(900)$ and ${\bar \kappa}(900)$,  which
are not quite established experimentally.
In  1998 edition of Particle Date Group \cite{PDG}, the $\sigma(560)$ is
listed as  $f_0(400-1200)$ with a very uncertain properties; a mass
between 400  to 1200 MeV and a very broad decay width  between 600 to 1000
MeV.  In a recent  experimental study of $\tau $ lepton decay by CLEO
collaboration \cite{CLEO},  a significant contribution due to the $\sigma$
is pointed out and it is  reported that inclusion of a  $\sigma$ with
$m_\sigma = 555$ MeV and $\Gamma_\sigma = 540$ MeV significantly improves
the fits.  The situation of $\kappa (900)$ is not clear experimentally.
Therefore,  altogether there are  9 possible candidates for lowest lying
scalar mesons.

The situation of these scalars is also not clear from the theoretical
point
of view;  there are model dependent calculations,  and as a  result,
different conclusions.   However,  within any theoretical framework 
two basic questions should be addressed:

1. Is there a clear evidence for the existence of a $\sigma (560)$ and 
a $\kappa (900)$?

2. Can we describe the properties of these mesons like their
masses, decay widths, interactions, and  in particular their quark
substructure?

In this talk we first review the general nonlinear chiral Lagrangian
framework of 
refs. \cite{Blk98,Blk99,Far99}, and discuss how within this framework 
a  $\sigma(560)$ was observed in \cite{San95,Har96}, and a $\kappa(900)$
was observed in \cite{Blk98}.
We then review how this nonlinear chiral Lagrangian can be rewritten in
terms of a scalar nonet by introducing a few new free
parameters.  We show how these parameters as well as the
acceptable range of the scalar mixing angle can be fixed by 
considering the $\pi\pi$ and  $\pi K$ scattering, and the
$\eta'\rightarrow\eta\pi\pi$  decay.   Based on the acceptable range of
the scalar mixing angle we
discuss that the quark substructure of the light scalar mesons is closer
to a four quark picture. We conclude by summarizing the results.
\section{ Our Theoretical Framework and Evidence for the $\sigma(560)$
and $\kappa(900)$}
We work within the effective non-linear chiral Lagrangian framework. The
pseudoscalar part of the Lagrangian is \cite{Blk98}
\begin{equation}
{\cal L}_\phi=
- { {F_\pi^2} \over 8} {\rm Tr}
\left( \partial_\mu U \partial_\mu U^\dagger  \right)
+{\rm Tr}
\left[
       {\cal B} \left( U + U^\dagger \right)
\right]
\label{L_phi}
\end{equation}
with $F_\pi = 131$ MeV,  ${\cal B}=m_\pi^2F_\pi^2/8 {\rm diag} (1,1,2
m_K^2/m_\pi^2-1)$ is the dominant symmetry breaking term, and 
$U=e^{2i {\phi\over F_\pi} }=\xi^2$ where
$\phi$ is the pseudoscalar nonet
\begin{equation}
\phi_a^b=
\left[ \begin{array}{c c c}
{ {\eta_{NS} + \pi_0^0}\over \sqrt{2}} &\pi^+&K ^+\\
\pi^-&{ {\eta_{NS} - \pi_0^0}\over \sqrt{2}}   &K ^0\\
K^-&{\bar K}^0&\eta_S
\end{array} \right].
\label{nonet_phi}
\end{equation}
$U$ transforms linearly under chiral transformation
[$U\rightarrow U_L U U_R^\dagger$ with 
$U_{L,R}\in U(3)_{L,R}$], whereas  $\xi$ transforms
nonlinearly [$
\xi\rightarrow U_L\xi K^\dagger(\phi, U_L, U_R)=K(\phi, U_L,
U_R)\xi U_R^\dagger
$].

The vectors can be introduced in this framework in terms of the vector 
nonet $\rho$ with a Lagrangian 
that has the same form as that of usual gauge fields
\begin{equation}
{\cal L}_\rho =
-\frac{1}{2} m_v^2 \mbox{Tr}
\left[ \left(\rho_\mu - \frac {v_\mu}{\widetilde{g}} \right)^2 \right]
 - \frac{1}{4} \mbox{Tr}
\left[ F_{\mu\nu}(\rho) F_{\mu\nu}(\rho) \right]
\label{L_rho}
\end{equation}
with $F_{\mu\nu} = \partial_\mu \rho_\nu -
\partial_\nu \rho_\mu - i \widetilde{g}
[ \rho_\mu , \rho_\nu ]$.  In (\ref{L_rho}),  
$
p_\mu = \frac{i}{2} 
\left(
  \xi \partial_\mu \xi^{\dag} - \xi^{\dag} \partial_\mu \xi
\right)
$
and
$
v_\mu = \frac{i}{2}
\left(
  \xi \partial_\mu \xi^{\dag} + \xi^{\dag} \partial_\mu \xi
\right) 
$
, and  have simple transformation properties under chiral transformation.

We introduce scalars into this picture in two stages.  First in order
to see whether there is an indication of $\sigma (560)$ and $\kappa(900)$
within our framework, we introduce scalars in a phenomenological way.
We consider a general isospin invariant form \cite{Blk98}
\begin{eqnarray}
{\cal L}_s &=& -\frac{\gamma_{\sigma \pi \pi}}{\sqrt{2}}
\sigma \partial_{\mu} \mbox{\boldmath ${\pi}$}\cdot
\partial_{\mu}{\mbox{\boldmath ${\pi}$}} -
\frac{\gamma_{\sigma K \bar K}}{\sqrt{2}} \sigma \left(
\partial_{\mu}K^{+} \partial_{\mu}K^{-} + \cdots  \right)  
-\frac{\gamma_{f_0\pi \pi}}{\sqrt{2}}
f_0 \partial_{\mu} \mbox{\boldmath ${\pi}$}  \cdot \partial_{\mu}
\mbox{\boldmath ${\pi}$} 
\nonumber \\
&&-
\frac{\gamma_{f_0 K\bar K}}{\sqrt{2}} f_0 \left(
\partial_{\mu}K^{+} \partial_{\mu}K^{-} + \cdots \right) 
 - {\gamma_{\kappa K\pi}} \left(
\kappa^0\partial_{\mu} K^{-} \partial_{\mu} \pi^+ + {....} \right) 
\label{L_s1}
\end{eqnarray}  
and take the coupling
constants as independent parameters -- we either take the couplings as
fitting parameters or input them from experimental measurements.

Now in order to see whether this framework sees a $\sigma$ and/or
a $\kappa$, we can consider computing processes to which  these mesons
could significantly contribute,  and then compare our prediction to the
experimental data and search for signs of these scalars.  In principle
$\sigma(560)$ and $\kappa(900)$ can be probed in $\pi\pi$ and
$\pi K$ scattering,  respectively.  In fact,  their 
contributions could be substantial as they appear as poles in 
the scattering amplitudes. 
To see the effect of $\kappa$,  using Lagrangian (\ref{L_s1}),
appropriate $\pi K$ scattering amplitudes were computed in \cite{Blk98}
and the result were matched to the experimental data.
The scattering amplitudes were computed by only taking tree level Feynman
diagrams into account -- this is motivated by $1/N_c$ expansion. 
It was shown in \cite{Blk98}  that a $\kappa$ with a mass around
900  MeV and a decay width around 320 MeV is needed in order to describe
the experimental data on the $\pi K$ scattering.   
This technique was first developed in \cite{San95,Har96}, in which
it was shown that in order to
agree with the $\pi\pi$ experimental  data there is a need for a $\sigma$
meson with a mass around 550 MeV and a decay width around 370 MeV.

Therefore within our theoretical framework there is a clear signal for
both the $\sigma(560)$ and the $\kappa(900)$.  
This motivates us, in our second stage of investigation,  to combine these
scalars together with the $f_0(980)$
and  the $a_0(980)$ into a light scalar nonet,  in terms of which we
rewrite our
Lagrangian in the next section.
\section{A Possible Scalar Nonet Below 1 GeV}
We now construct a light scalar nonet out of
$\sigma(560)$, $\kappa(900)$,  $f_0(980)$ and $a_0(980)$ in the form
\begin{equation}
N = \left[ \begin{array}{c c c}
N_1^1&a_0^+&\kappa ^+\\
a_0^-&N_2^2&\kappa ^0\\
\kappa^-&{\bar \kappa}^0&N_3^3
\end{array} \right].
\label{nonet_N}
\end{equation}
In general, $\sigma(560)$ and $f_0(980)$ are a mixture of
$(N_1^1+N_2^2)/\sqrt {2}$ and  $N_3^3$.  We represent this mixing in terms
of a scalar mixing angle $\theta_s$ as
\begin{equation}
\left( \begin{array}{c} \sigma\\ f_0 \end{array} \right) = \left(
\begin{array}{c c} {\rm cos} \theta_s & -{\rm sin} \theta_s \\ {\rm sin}
\theta_s & {\rm cos} \theta_s \end{array} \right) \left( \begin{array}{c}
N_3^3 \\ \frac {N_1^1 + N_2^2}{\sqrt 2} \end{array} \right).
\label{sf_mix}
\end{equation}
In this parametrization, $\theta_s = \pm\pi/2$ corresponds to  the
conventional ideal mixing when a pure $q\bar q$ assignment
is used for $N$.  Another interesting limit is  $\theta_s=0$ which
corresponds to the
dual ideal mixing when a pure four-quark assignment is used to describe
$N$ \cite{Jaf77}.

\begin{figure}[b]
\begin{center}
\epsfig{file=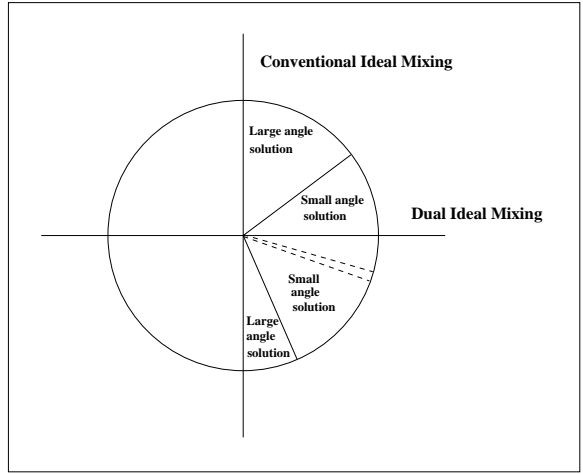,width=3in}
\end{center}
\vspace{10pt}
\caption{
Two regions for the scalar mixing angle for the acceptable range of 
685 MeV$\le m_\kappa \le$ 980 MeV.  As $m_\kappa$ varies  from its
minimum value to its maximum value, $\theta_s$ in the small angle region 
varies from 36$^o$ to -71$^o$, and in the large angle region
$\theta_s$ varies
from 36$^o$ to 90$^o$, then to -90$^o$ and finally to -71$^o$.  The region
bounded
between dashed lines ($-20^o\le m_\kappa \le -15^o$) corresponds to 875
MeV
$\le m_\kappa \le$ 897 MeV, and
is consistent with experimental data on $\pi\pi$ and $\pi K$
scattering,
as well as on $\Gamma [f_0(980)\rightarrow\pi\pi]$. }
\label{Fig_theta_s}
\end{figure}

We can now rewrite the scalar sector of our Lagrangian in terms of the
nonet (\ref{nonet_N})
\begin{eqnarray}
{\cal L}_{mass} &=&-a {\rm Tr}(NN) - b {\rm Tr}(NN{\cal M}) - c {\rm
Tr}(N){\rm Tr}(N) 
- d {\rm Tr}(N) {\rm Tr}(N{\cal M})
\nonumber \\
{\cal L}_{N\phi \phi} &=&
A{\epsilon}^{abc}{\epsilon}_{def}N_{a}^{d}{\partial_\mu}{\phi}_{b}^{e}
{\partial_\mu}{\phi}_{c}^{f}
+ B {\rm Tr} \left( N \right) {\rm Tr} \left({\partial_\mu}\phi
{\partial_\mu}\phi \right)  
\nonumber \\
&&+ C {\rm Tr} \left( N {\partial_\mu}\phi
\right) {\rm Tr} \left( {\partial_\mu}\phi \right) 
+ D {\rm Tr} \left( N \right) {\rm Tr}
\left({\partial_\mu}\phi \right)  {\rm Tr} \left( {\partial_\mu}\phi
\right)
\label{Npp}
\end{eqnarray}
with ${\cal M}=$diag $(1,1,x)$ is the spurion matrix with $x$ the
ratio of strange to  non-strange quark masses.  The mass part of the
Lagrangian is given 
in terms of new free parameters $a,b,c,d$, and $\theta_s$ which can be
determined by inputting the scalar masses  $m_\sigma, m_{f0}, m_\kappa$
and $ m_{a0}$.  
We find that our model 
restricts $m_\kappa$ in the range 685 to 980 MeV, and also for any
input of scalar masses there are two solutions for $\theta_s$. 
Thus,  in our framework there are two acceptable ranges for the
scalar mixing angle which are shown in Fig. \ref{Fig_theta_s}.
These are  the {\it large angle solution}: 
$36^o \le \theta_s \le 90^o$ and $-90^o \le \theta_s \le -71^o$, and the
{\it small angle solution}: $-71^o \le\theta_s \le 36^o$.  
As $m_\kappa$ varies from its minimum value to its maximum, these 
regions are entirely swept through.
\begin{figure}[b]
\begin{center}
\epsfig{file=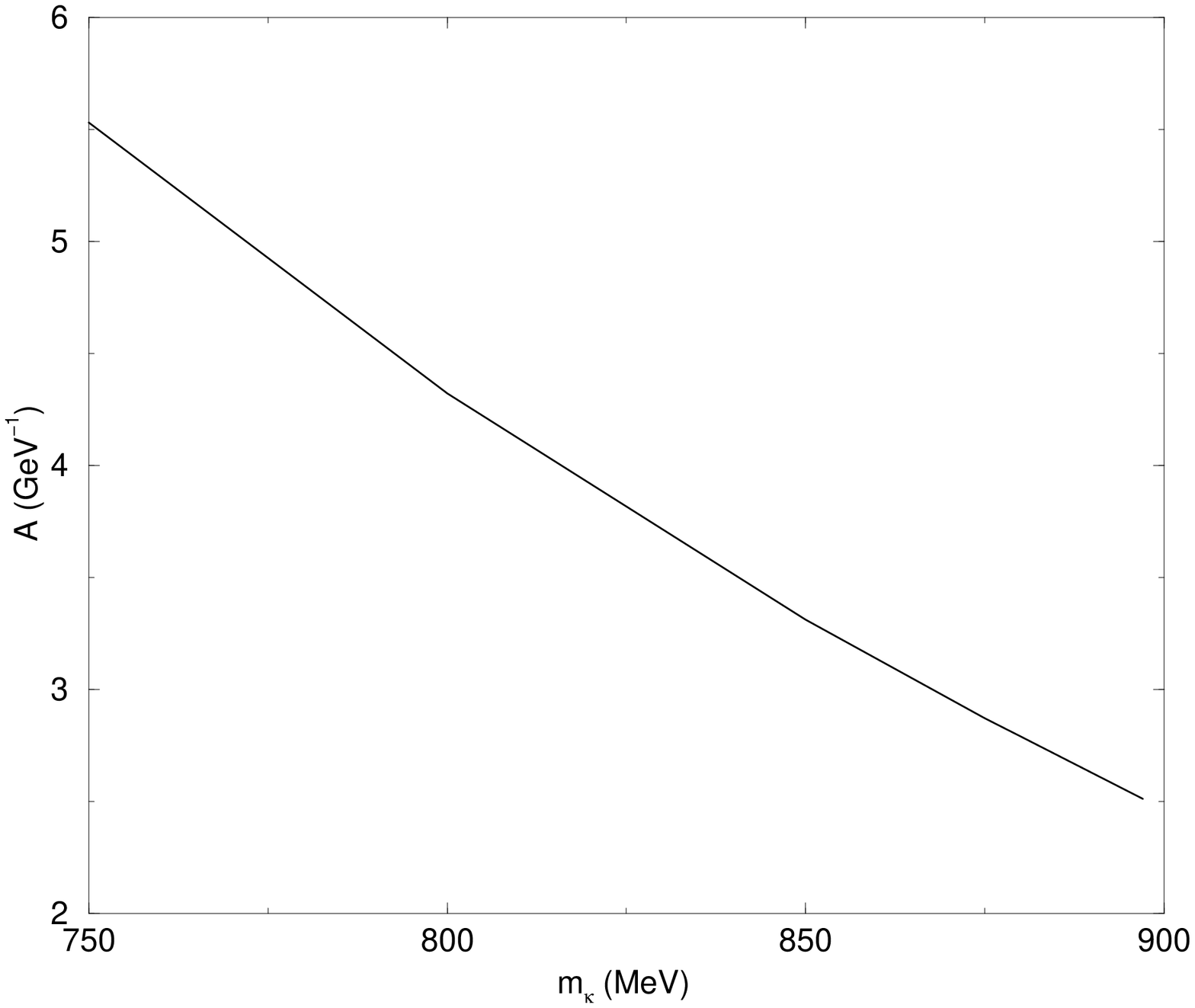,width=2.3in, angle=270}
\epsfig{file=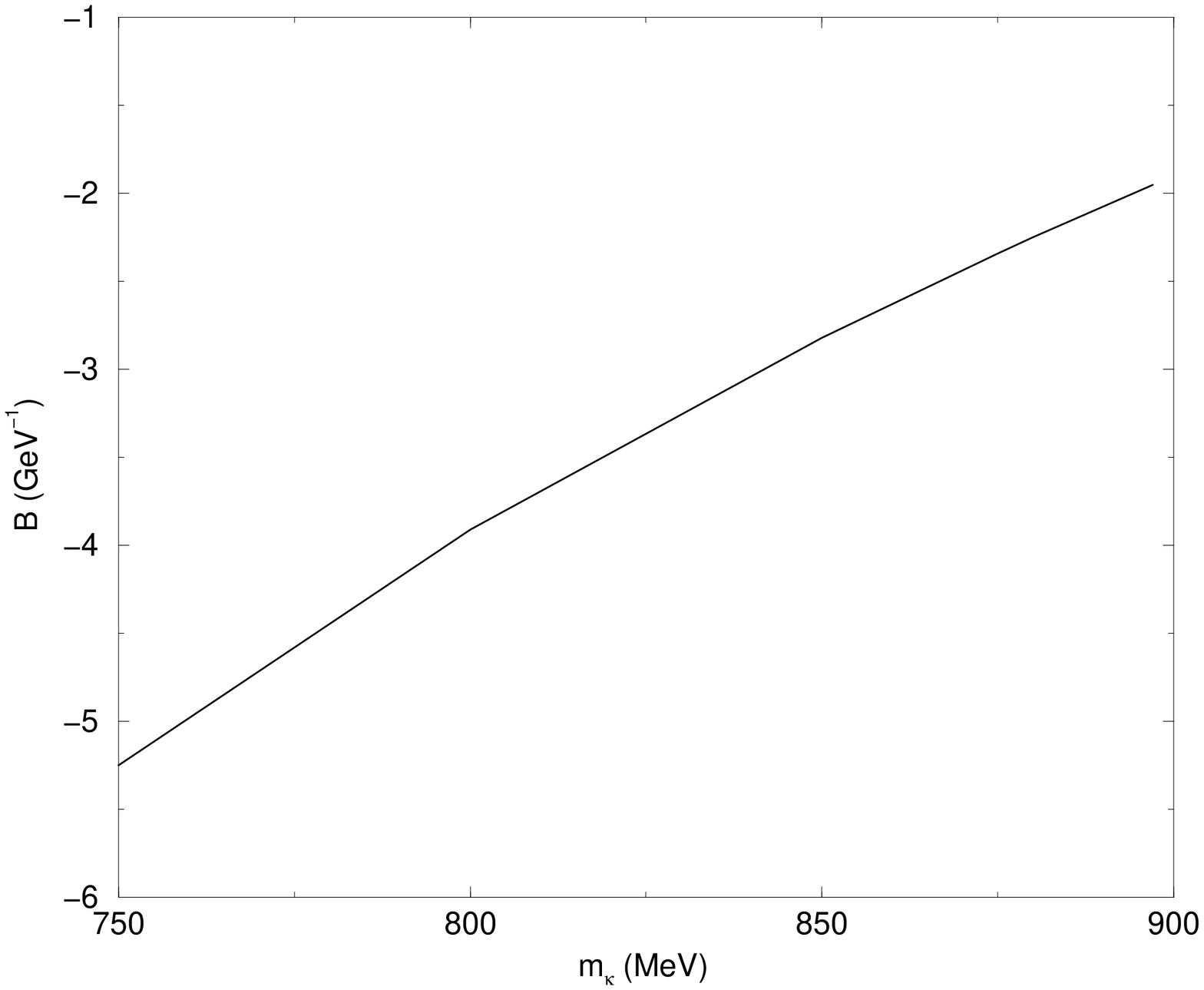,width=2.3in, angle=270}

\epsfig{file=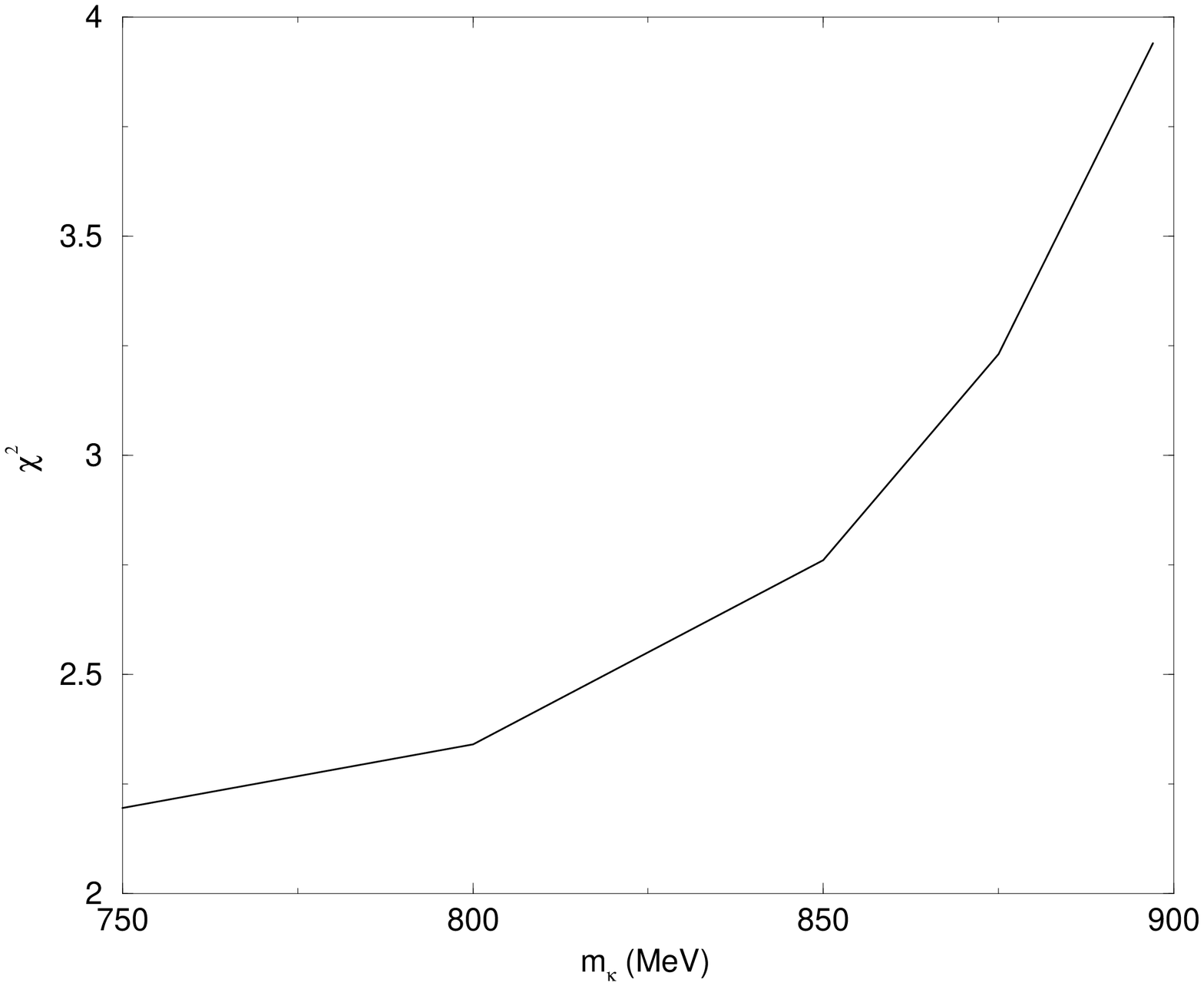,width=2.3in, angle=270}
\epsfig{file=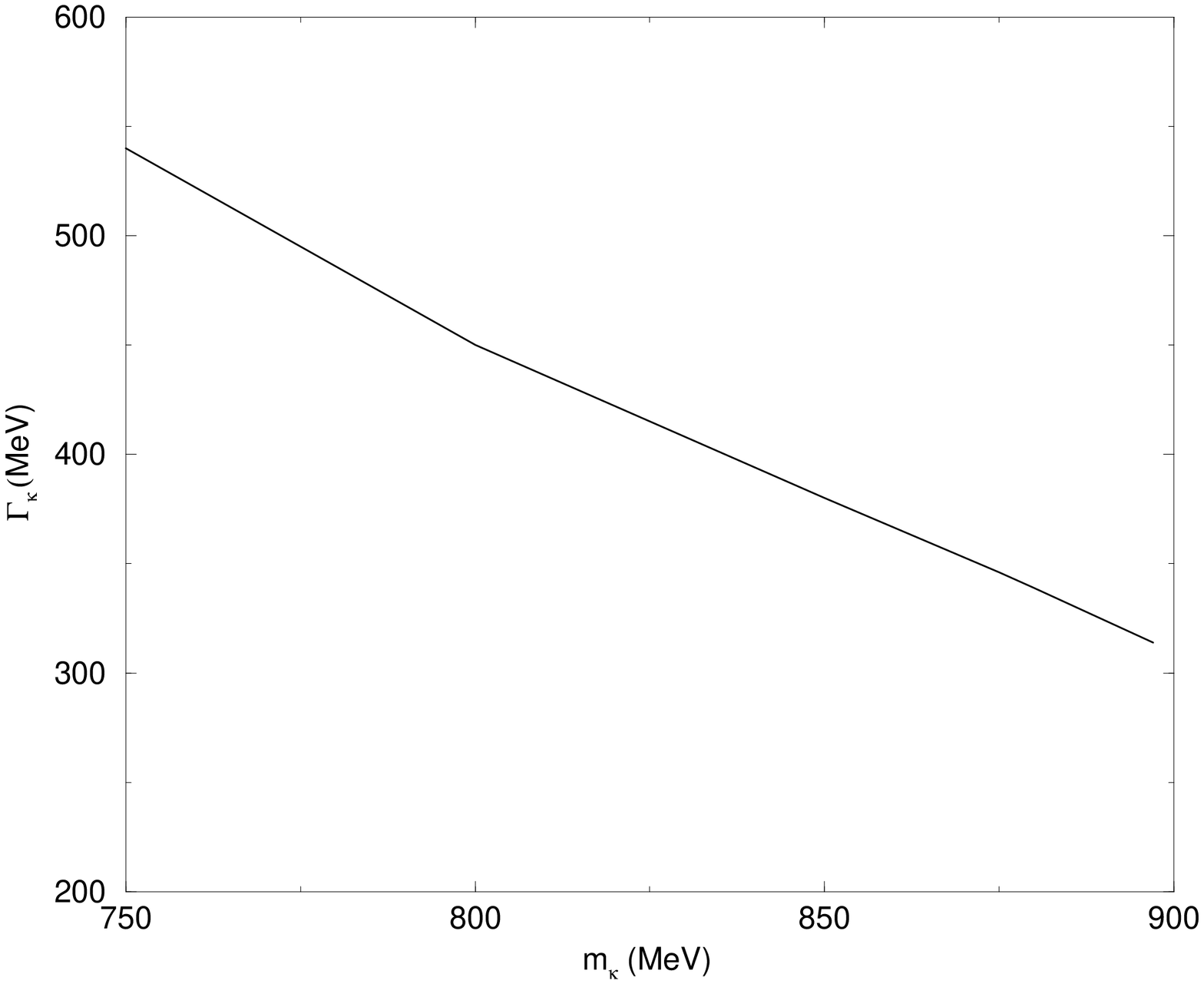,width=2.3in, angle=270}
\end{center}
\vspace{10pt}
\caption{
$m_\kappa$ dependence of the fitting parameters $A$, $B$ and
$\Gamma_\kappa$,  in a fit of the theoretical prediction  of the $\pi K$
scattering amplitude to the experimental data,  together with the
$\chi^2$ of the fit.
}
\label{Fig_mk}
\end{figure}
\begin{figure}[t]
\begin{center}
\epsfig{file=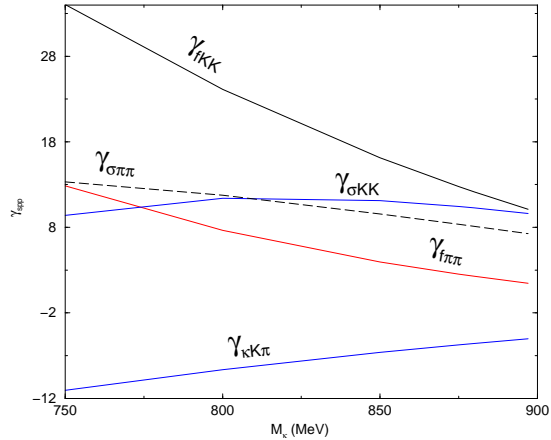,width=2.3in, angle=270}
\end{center}
\vspace{10pt}
\caption{
$m_\kappa$ dependence of the hadronic couplings found in a 
fit of the theoretical prediction  of the $\pi K$
scattering amplitude to the experimental data.
}
\label{Fig_gamma}
\end{figure}
There are also new free
parameters $A, B, C$, and  $D$
in the scalar-pseudoscalar-pseudoscalar interaction part of Lagrangian
which can be determined by appropriately matching our theoretical
prediction to the experimental data.  
A consequence of introducing the nonet (\ref{nonet_N}) is that the
scalar  couplings are now related to each other by the underlying chiral
symmetry, i.e. $\gamma_{spp} = \gamma_{spp} ( A, B, C, D, \theta_s,
\theta_p )$, where $\theta_p$ is the pseudoscalar mixing angle 
for which we choose the value $37^o$.

We numerically search through both ranges of $\theta_s$ and fit our
prediction for $\pi K$ scattering amplitude
to the experimental data.   This determines $A$ and $B$ in the interaction
Lagrangian.
We find that the $\chi^2$ of fit improves as 
we lower $m_\kappa$.  This is shown in Fig. \ref{Fig_mk}, together with
the
$m_\kappa$ dependence of the parameters $A$ and $B$ in  the interaction
Lagrangian, and the total decay width of $\kappa$.   Although
the $\chi^2$ fit improves for lower values of $m_\kappa$,  other
experimental 
data further restricts the acceptable range of $m_\kappa$.
We take into account limits from
$\pi\pi$ scattering amplitude \cite{San95,Har96} on the
strong interaction couplings $\gamma_{spp}$,  as well as the decay
width $\Gamma [f_0(980)\rightarrow\pi\pi]$.  
\hskip .3cm The $m_\kappa$ dependence of
the resulting couplings are shown in Fig. \ref{Fig_gamma}.  We find that
the small
angle solution is favored as it contains a small region 
(corresponding to 875 MeV $\le m_\kappa \le$ 897 MeV) that is consistent
with these experimental constraints.  This region, which is bounded by
dashed lines in
Fig. \ref{Fig_theta_s}, is obviously close to $\theta_s = 0$.  This is 
how our model indirectly  probes the quark substructure of this scalar
nonet
-- the fact that $\theta_s$ is small  means that the scalar
mixing  in our model is closer to the dual ideal mixing and therefore a
four-quark scenario is favored for this nonet.
\begin{figure}[t]
\begin{center}
\epsfig{file= 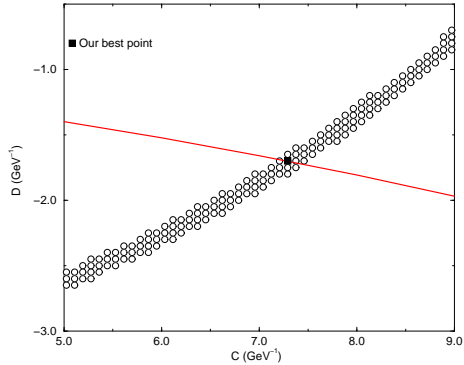, width=2.3in, angle=270}
\end{center}
\vspace{10pt}
\caption{
Regions in $CD$ plane consistent with two different experimental 
measurements on $\eta'$ decay. 
Regions consistent with $\Gamma^{exp.} [\eta'\rightarrow\eta\pi\pi ]$
are represented by circles.
The solid line represents points consistent with the energy 
dependence of the normalized magnitude of the decay matrix element.
}
\label{Fig_CD}
\end{figure}
\begin{figure}[t]
\begin{center}
\epsfig{file=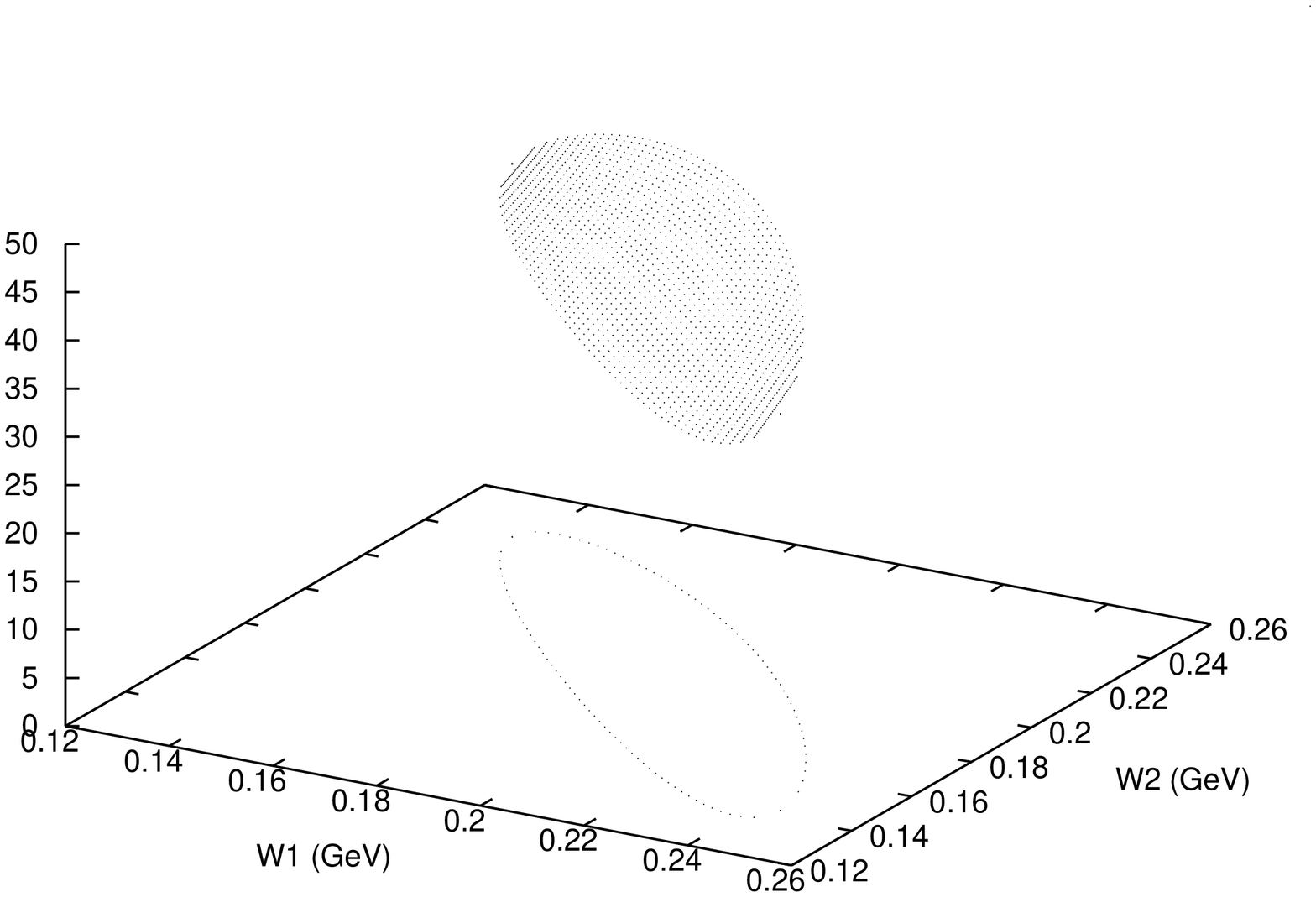,width=2.3in, angle=270}
\end{center}
\vspace{10pt}
\caption{
Energy dependence of the magnitude of the $\eta'\rightarrow\eta\pi\pi$
decay matrix  element.   $w_1$ and $w_2$ are the total energy of the final
state pions,  and are bounded within the ellipse-like region in the
$\omega_1\omega_2$ plane.
}
\label{Fig_Energy}
\end{figure}
\section{$\eta'\rightarrow\eta\pi\pi$ Decay}
In the previous section we rewrote  the scalar part of the Lagrangian in
terms
of a scalar nonet $N$.   We evaluated all free parameters in the
Lagrangian except $C$ and $D$ in the interaction piece in (\ref{Npp}).
These
two parameters were probed in ref. \cite{Far99} by matching the prediction
of
our model for the partial decay width of $\eta'\rightarrow\eta\pi\pi$, and 
for the energy dependence of the normalized magnitude of the decay
matrix element,  to the experimental data.  The same values 
of $A$, $B$ and $\theta_s$ that were found in ref. \cite{Blk99} were used
in
this decay analysis.   The $CD$ parameter space was scanned numerically
and was searched for the physical regions that describe both
experimental measurements of this decay.  The result
is  shown in Fig. \ref{Fig_CD}.   The gray region is consistent with the
partial decay width of this decay, and the solid line is consistent with
the energy dependence of the normalized magnitude of the decay matrix
element.  Their intersection in the $CD$  plane ($C\approx 7.3$
GeV$^{-1}$ and
 $D\approx -1.7$ GeV$^{-1}$) {\it exists} and is {\it unique}.  This means
that there
is a unique choice of  free parameters of Lagrangian (\ref{Npp}) that,  in addition to
$\pi\pi$  and $\pi K$ scattering amplitudes, describes the
$\eta'\rightarrow\eta\pi\pi$ decay.  The energy dependence of this decay 
is plotted in Fig. \ref{Fig_Energy}.

As a by-product we compute,  with the same extracted $C$ and $D$,
the  partial decay width of $a_0(980)\rightarrow\pi\eta$ to be
approximately 65 MeV.  This,  together with the
$\Gamma [a_0(980)\rightarrow K {\bar K} ]\approx 5 MeV$ found in
\cite{Blk99},
provide an estimate of the total decay width of $a_0(980)$ around 70 MeV.
This is in a very close agreement with a recent experimental analysis of
$a_0(980)$ in ref. \cite{Tei99}.  
\section{Summary and Discussion}
In this talk we reviewed the light scalar mesons
in the non-linear  chiral Lagrangian framework of references
\cite{Blk98,Blk99,Far99}.
We saw that in this approach there is a need for the  $\sigma(560)$ and
the $\kappa(900)$ in order to be able to describe the experimental data on
the $\pi\pi$ and $\pi K$ scattering. We then constructed a light scalar
nonet below 1 GeV consisting of $\sigma(560)$, $\kappa(900)$, $f_0(980)$,
and
$a_0(980)$,  and rewrote the Lagrangian in terms of this nonet by
introducing eight new free parameters.
We showed that with these parameters we can describe many experimental
facts including the scalar mass spectrum,  their interactions
with pseudoscalars in $\pi\pi$ and $\pi K$ scattering,  the $\eta'$
decay, and the decay width of $f_0(980)$.  We could predict the
total decay width of $a_0(980) \approx 70$ MeV in a very close
agreement with a recent experiment \cite{Tei99}.
We discussed that although the chiral
Lagrangian model presented here is entirely formulated in terms of the   
meson fields and in principle does not know anything about the underlying
quark  substructure,  in practice,  the knowledge of the mixing angle
indirectly probes the quark substructure of
these scalars.   We saw,  through a careful numerical analysis, that the
acceptable range of the mixing angle is such that it suggests the quark
substructure of these scalars is closer to a four quark picture.  This is
in agreement with a recent theoretical investigation \cite{Ach99}.
\acknowledgments
This talk was based on works done in collaboration with D. Black, F.
Sannino and J. Schechter \cite{Blk98,Blk99,Far99},  and  has been
supported in part by DE-FG-02-92ER-40704. We also thank the
organizers of the MRST'99 for a very interesting conference.


\begin{references}
\bibitem{Blk98}Black, D.,  Fariborz, A.H., Sannino, F., and 
Schechter, J., 
{\it Phys. Rev. D} {\bf 58}, 054012-1-11 (1998). 
\bibitem{Blk99}Black, D.,  Fariborz, A.H., Sannino, F., and 
Schechter, J., {\it Phys. Rev. D} {\bf 59}, 074026-1-12 (1999).
\bibitem{Far99}Fariborz, A.H.,  and Schechter, J., 
{\it Phys. Rev. D}, {\bf 60}, 034002-1-11 (1999).
\bibitem{PDG}Review of Particle Physics, {\it Euro. Phys. J. C} {\bf
3} (1999).
\bibitem{CLEO} 
Asner, D.M.,  et al, CLEO Collaboration, hep-ex/9902022. 
\bibitem{San95}
Sannino, F., and Schechter, J., {\it Phys. Rev. D}  {\bf 52},  96-107
(1995).
\bibitem{Har96}
Harada, M., Sannino, F.,  and Schechter, J., {\it Phys. Rev. D}
{\bf 54}, 1991-2005 (1996); {\it Phys. Rev. Lett. } {\bf 78}, 1603
(1997).
\bibitem{Jaf77}Jaffe, R.L., {\it Phys. Rev. D} {\bf 15}, 267-289 
(1977).
\bibitem{Tei99}Teige, S.,  et al, {\it Phys. Rev. D} {\bf 54}, 
012001-1-12 (1999).
\bibitem{Ach99}
Achasov, N.N.,  {\it Phys. Usp.} {\bf 41}, 1149-1153 
(1998); 
Achasov, N.N.,  and Shestakov, G.N.,  
hep-ph/9904254.
\end{references}
\end{document}